\documentclass[prd,aps,nofootinbib,preprint]{revtex4}
\usepackage{amsmath,amssymb,amsfonts,color,graphicx,graphics,latexsym,placeins,epsfig}
\usepackage{footmisc}
\usepackage[compatibility=false]{caption}
\usepackage{subcaption}
\usepackage{bbold}
\usepackage{enumitem}
\usepackage{hyperref}
\DeclareMathOperator{\sech}{sech}
\usepackage{natbib}
\usepackage{float}
\newcommand{\be}{\begin{equation}}
\newcommand{\ee}{\end{equation}}
\newcommand{\ba}{\begin{eqnarray}}
\newcommand{\ea}{\end{eqnarray}}

\newcommand*{\myprime}{^{\prime}\mkern-1.2mu}
\newcommand*{\mydprime}{^{\prime\prime}\mkern-1.2mu}

\newcommand*{\id}{{\rm\hbox{1\kern-0.15em \vrule width .1pt depth-.2pt}}}
\renewcommand{\inf}{\infty}
\begin{document}


\title{\Large \bf Geodesic congruences in exact plane wave spacetimes and the 
memory effect}
\author{Indranil Chakraborty${}^{1}$ and
Sayan Kar ${}^{1,2}$}
\email{indradeb@iitkgp.ac.in, sayan@phy.iitkgp.ac.in}
\affiliation{${}^1$ Centre for Theoretical Studies \\ Indian Institute of Technology Kharagpur, Kharagpur 721 302, India}
\affiliation{${}^{2}$ Department of Physics \\ Indian Institute of Technology Kharagpur, Kharagpur 721 302, India.}

\begin{abstract}
\noindent  
Displacement and velocity memory effects
in the exact, vacuum, plane gravitational wave line element have been studied
recently by looking at the behavior of pairs of geodesics or via geodesic deviation.
Instead, one may investigate the evolution of geodesic congruences.
In our work here, we obtain the evolution of the kinematic variables 
which characterize timelike geodesic congruences, using chosen pulse profiles (square and sech squared) in the exact, plane gravitational wave line element. We also analyse the behavior of
geodesic congruences in possible physical scenarios describable using derivatives (first, second and third) of one of the chosen pulses.
Beginning with a discussion on the generic behaviour of such congruences and consequences thereof, we find exact analytical expressions for shear and expansion with the two chosen pulse profiles. Qualitatively similar numerical
results are noted when various derivatives of the sech-squared pulse are used.
We conclude that for geodesic congruences, a growth (or decay) 
of shear causes focusing of an initially parallel congruence, after the departure of the pulse. A correlation between the "focusing time" (or $u$ value, $u$ being the affine parameter) and the amplitude of the pulse (or its derivatives) is found. 
Such features distinctly suggest a memory effect, named in
recent literature as ${\cal B}$ memory.
\end{abstract}


\maketitle

\section{Introduction}

\noindent The memory effect in gravitational wave physics has been a topic of
active research interest in recent times \cite{memoryreview,lisa}. Though yet to be observed in
gravitational wave detectors there have been proposals \cite{lisa,aligoproposals} about how it can be seen
in advanced versions of the present-day detectors. The physics 
of memory is related to a net displacement (or a residual velocity) 
noted in freely falling detectors, caused by the passage of a pulse of 
gravitational radiation. This leads to a permanent change in 
the Minkowski spacetimes that exist before the arrival of the pulse and after
its departure. The change is connected with spacetime diffeomorphisms taking one asymptotically flat spacetime to another, which do not tend to identity at infinity.  Asymptotically flat spacetimes before the arrival of the pulse and after its departure are therefore inequivalent. It is known that they may be related via Bondi-Metzner-Sachs transformations (eg. super-translations) \cite{bms,bms1}.        

\noindent The first report of such an effect appears in
the context of gravitational collapse in globular clusters as noted by Zel'dovich and Polnarev \cite{zeld}. Subsequently, Braginsky and Grishchuk \cite{brag}, while working within linearized gravity, defined the memory effect
to be the difference between the quadrupole moments of the source at initial and  final times. Later, Christodoulou \cite{christ} showed that there is a nonlinear contribution to the effect and argued that this is due to the effective stress energy of the gravitational waves transported to null infinity. Thorne \cite{thorne} had argued that the nonlinear contribution to Christodoulou memory could be attributed to gravitons sourced by a gravitational wave burst. Following this idea, Bieri {\em et al}. \cite{bieri1,wald} demonstrated a contribution to memory from other particles having zero rest mass. Thus, they were able to
distinguish linear and nonlinear contributions as ordinary and null memory. The stress energy travels to null infinity in the latter case only. Apart from four-dimensional asymptotically flat spacetimes, there exists work on the memory effect in other spacetimes and in higher dimensions  \cite{bieri1,wald,hamada}. 
Memory effects have also been discussed 
in the context of modified gravity and massive gravity theories \cite{modifiedgrav}.

\noindent Very recently, Zhang {\em et al.} \cite{zhang,zhang1,zhang2} have tried to arrive at the memory effect in the well-known exact plane gravitational wave spacetimes
\cite{plane,brink,BJR}. Apart from other analyses in their paper \cite{zhang}, they studied geodesics in this geometry by assuming certain specific forms of the functions which appear in the line element. In particular, they chose a
Gaussian pulse (and its derivatives) and numerically solved the geodesic equations 
to obtain some
qualitative results on the displacement and velocity memory effects. The appearance of a net relative displacement and/or a net relative velocity caused by the passage of a pulse are termed as the displacement and velocity memory effects respectively.
Building on these ideas
we shall show in our work how the behaviour of geodesic congruences may also lead to a memory effect via a change in the shear
and expansion of the congruence, caused by the 
pulse. This memory effect involving
geodesic congruences is closer to velocity
memory but not quite the same.

\noindent In general, a memory effect can 
therefore be arrived at
in three different ways, eventually
leading to qualitatively similar broad conclusions. Let us now briefly discuss
each and note the differences
between them too. We will confine ourselves
to the sandwich pulse profiles in
exact plane wave metrics \cite{bondi,plane} 
while studying memory effects. 
From a motivational standpoint, one may argue that any pulse observed in a future detection of a binary merger event is likely to be finite for a certain range of $u$. The Fourier decomposition of such signals would exhibit a peak frequency (chirp) of the burst itself over a quasistatic low frequency background as observed in the detection events by LIGO \cite{LIGO}. However, it goes without
saying, that the exact plane wave metric
is largely theoretical and has no direct link
with present-day gravitational wave observations.

\noindent The first among the three
ways involves 
obtaining a
{\em net displacement}
between pairs of geodesics, caused by a gravitational wave pulse, after the pulse has left. This may be found by {\em directly integrating the geodesic equation} to
obtain the evolution of the separation of
each coordinate, for pairs of geodesics.
A second way, largely related to the previous one, is to
{\em directly integrate the geodesic deviation equation} and understand the evolution of the 
deviation vector. Both these approaches are associated with
displacement memory.  Additionally, 
the former may be used to arrive at a velocity memory as discussed in \cite{zhang,zhang1,zhang2}. 

\noindent Here we choose the third way of arriving at a memory effect,
namely by looking at the behavior of geodesic congruences. This approach is 
covariant and has been proposed in a recent article by O'Loughlin and Demirchian
\cite{congruence} wherein the
term ${{\cal B}}$-memory (${\cal B}$ denotes the tensor ${\cal B}^i_{\,\,j}$, the covariant gradient of the velocity field) is introduced in the context of impulsive gravitational
waves. Our work supports and extends the
proposal in \cite{congruence} using the simplest class of {\em pp}-wave spacetimes--the well-known exact plane gravitational waves.

\noindent In the exact plane gravitational
wave spacetime, there arise
free profile functions [$A_{+}(u)$ or $A_\times (u)$]. Apart from generic results
obtained without choosing specific functional forms for the profiles, 
we also find exact results with simple pulse profiles (e.g., a {\em square pulse} and a 
{\em sech-squared pulse}). 
Further, for the sech-squared pulse we analyze memory using the first, second and third derivatives (which may arise
in different physical contexts) of the pulse. 
\footnote{
In a recent paper
\cite{shore}, Shore has looked at the square pulse briefly in an Aichelburg-Sexl impulsive gravitational wave line element, which is different from the spacetime 
we work with here.}

\noindent
As is well known, a timelike geodesic congruence is  studied through the 
behavior of the expansion, shear and rotation which is comprised of trace, symmetric traceless and anti-symmetric parts of the ${\cal B}$-tensor. The nature of evolution of these kinematical variables 
associated with the congruence are first obtained qualitatively using simple inequality arguments. Subsequently, using the pulse profiles mentioned 
earlier, we  obtain the kinematic variables exactly or using numerical methods. By noting the point of convergence, we are able to relate the amplitude of the pulse with the {\em time}
at which it focuses. 
We demonstrate shear-induced focusing which causes a permanent change in the expansion after the departure of the pulse. This is ${\cal B}$-memory as introduced
in \cite{congruence}.
In other words, it is the expansion, shear and rotation which may
undergo a permanent change caused by the appearance of the pulse. 
Since the ${\cal B}$-tensor is the gradient of the normalized velocity field, ${\cal B}$-memory, as mentioned earlier, has a connection with {\em velocity memory}, 
though it is not quite the same.

\noindent An important feature of our work is analytical solvability. Unlike the Gaussian pulse or its derivatives, for our choices of the profiles, 
quite a bit can be done by exactly solving the Raychaudhuri equations. We explicitly 
illustrate the 
memory effect using the behavior of shear and expansion, for the chosen
pulse profiles, through our largely analytical results. However, the results for the derivatives of one of the pulses are numerically obtained.

\noindent In Sec. II, we write the line element of the 
vacuum, plane wave spacetimes in Brinkmann
coordinates and obtain the geodesic equations. As an illustration, we show the displacement and velocity
memory effects 
using a square pulse.
Section III deals with the evolution of the expansion, shear and rotation of geodesic congruences given by the Raychaudhuri equations.
A qualitative analysis for both the pulse and its derivatives 
is followed by exact solutions for the case of a pulse. 
In Sec. IV, we numerically analyze the physically interesting cases involving the derivatives of the 
continuous sech-squared pulse. Finally,
Sec. V is a summary of our results with some comments on future work.

\section{Geodesics in exact plane wave spacetimes}
\subsection{Brinkmann coordinates }
\noindent The exact plane wave spacetimes are a class among general {\em pp}-wave spacetimes which solve the vacuum Einstein field equations of general relativity \cite{brink,BJR,peres}. The metric components are the same at every point on each wave surface. The coordinate system employed in our calculation is the standard Brinkmann coordinates which are both harmonic and global. The line element in Brinkmann coordinates is given by the form
\begin{align}\label{eq:brinkmann}
ds^2=\delta_{ij}dx^idx^j+2dudV+K_{ij}(u)x^ix^jdu^2.
\end{align}

\noindent The gravitational field is encoded in the term $K_{ij}(u)$, which satisfies the wave equation
\begin{align}
    \Box \, ( K_{ij}(u)x^ix^j)=0.\label{eq:wave}
\end{align}
\noindent $K_{ij}(u)$ is a trace-free, $2\times2$ matrix having two independent components which are known as the polarizations of the plane gravitational wave (+,$\times$). We have
\begin{align}\label{eq:polarization}
K_{ij}(u)x^ix^j=\dfrac{1}{2}A_+(u)[x^2-y^2]+A_{\times}(u)xy.
\end{align}
\noindent The polarizations $A_+(u)$ (plus),   $A_{\times}(u)$ (cross) are functions of retarded time variable $u$.
\noindent Another coordinate system used for this metric is the  Baldwin-Jeffrey-Rosen (BJR) coordinate system \cite{rosen} which however suffers from the presence of coordinate singularities.



\subsection{The geodesic equations}

\noindent The geodesic equations in Brinkmann coordinates having both nonzero polarizations are given as
\begin{align}\label{eq:combined_x}
    \frac{d^2 x}{du^2}=\frac{1}{2}A_+(u)x+\frac{1}{2}A_\times(u)y,
\end{align}
\begin{align}\label{eq:combined_y}
    \frac{d^2 y}{du^2}=-\frac{1}{2}A_+(u)y+\frac{1}{2}A_\times(u)x,
\end{align}
\begin{align}\label{eq:combined_V}
   \frac{d^2 V}{du^2}+\frac{1}{4}\frac{dA_+(u)}{du}(x^2-y^2)+A_+(u)\Big(x\frac{dx}{du}-y\frac{dy}{du}\Big)+A_\times(u)\Big(y\frac{dx}{du}+x\frac{dy}{du}\Big)+\frac{1}{2}\frac{dA_\times(u)}{du}xy=0. 
\end{align}
\noindent Notice that we have used $u$ as an affine parameter. This is easily checked by writing down the Euler-Lagrange equation for the $V$ coordinate. The general form for $V(u)$ and $\dot{V}(u)$ is\footnote{In this paper, $\dot{f}=\frac{df}{du}$, for any general $f$. The two symbols are used interchangeably throughout the paper.} obtained by performing some algebra on Eq.(\ref{eq:combined_V}) and from the  geodesic Lagrangian
  (derived from the line element) in Eq.(\ref{eq:brinkmann}):
 \begin{align}\label{eq:dV} 
     \frac{dV}{du}=-\frac{1}{4}A_+\big(x^2-y^2\big)-\frac{1}{2}\bigg[\bigg(\frac{dx}{du}\bigg)^2+\bigg(\frac{dy}{du}\bigg)^2\bigg]-\frac{1}{2}A_\times(u)xy-\frac{k}{2},
 \end{align}
 
 \begin{align}
     V(u)=-\frac{1}{2}\bigg(x\frac{dx}{du}+y\frac{dy}{du}\bigg)-\frac{k}{2}u+C_1. \label{eq:V}
 \end{align}
 
 \noindent The solution for $V(u)$ contains the integration constant $C_1$ and also $k$, which is $0$ or $1$ for null or timelike geodesics, respectively. Thus, for any pulse of a given polarization, if Eqs.(\ref{eq:combined_x}) and (\ref{eq:combined_y}) for $x(u)$ and $y(u)$ are analytically solvable, then $V(u)$ also can be analytically obtained. Both the first and second integrals for the coordinate $V$ are known from $x$, $y$ and its derivatives.
Hence, Eq.(\ref{eq:combined_V}) reduces to an identity. 
The fact that coordinate $V$ does not give a new equation of motion is useful for the discussion in the next section where we analyze the kinematic variables (expansion, shear, rotation) associated with   
the velocity field on the two-dimensional transverse $xy$ plane 
with $u$ acting as a parameter ({\em similar to time in classical mechanics}).

 \noindent The geodesic Lagrangian for the exact plane gravitational wave line element written in Brinkmann coordinates, with $u$ as the affine parameter, is given as
\begin{align}
    \mathcal{L}=\dot{x}^2+\dot{y}^2+\frac{1}{2}A_{+}(u)(x^2-y^2)+A_{\times}(u)xy+2\dot{V}.\label{eq:Lagrangian}
\end{align}
\noindent It is clear that the last term on the rhs of $\mathcal L$ is a total derivative and hence has no effect on equations of motion for the "generalised" coordinates $x$ and $y$. The system becomes two dimensional for the parameter $u$ where the terms quadratic in $x,y$ gives the associated potential . The
resemblance with a two-dimensional system where the polarization factors $A_+(u)$ and $A_\times(u)$ act as time ($u$)-dependent squared frequencies of an oscillator
(or an inverted oscillator) and a time ($u$)-dependent $x-y$ coupling coefficient respectively, is clearly visible.
 
\subsection{Memory effects}

\begin{figure}[H]
 \centering
  \begin{subfigure}[t]{0.45\textwidth}
   \centering
   \includegraphics[width=\textwidth]{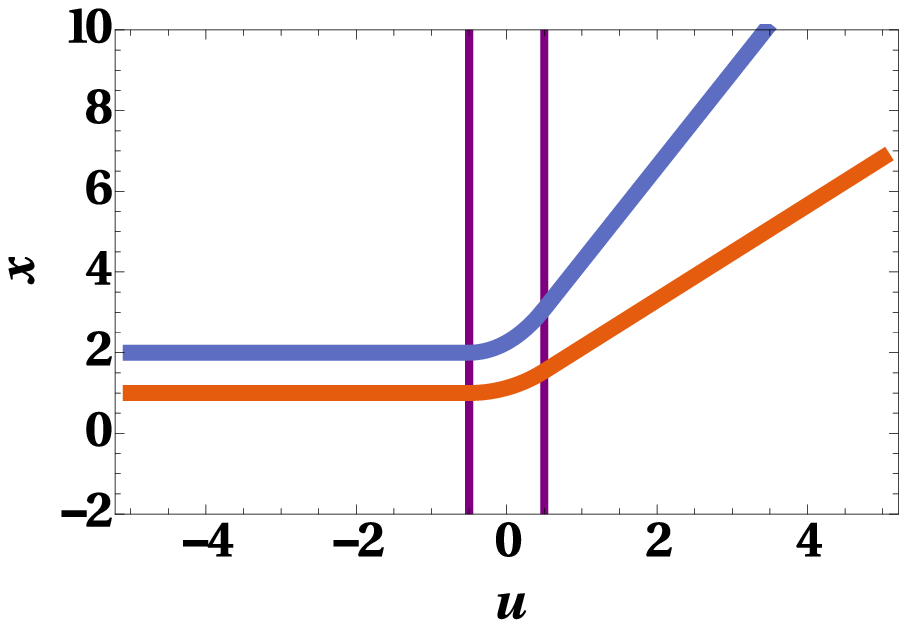}
   \caption{}
   \label{fig:x_plus}
  \end{subfigure}\hspace{1.2cm}
  \begin{subfigure}[t]{0.45\textwidth}
   \centering
   \includegraphics[width=\textwidth]{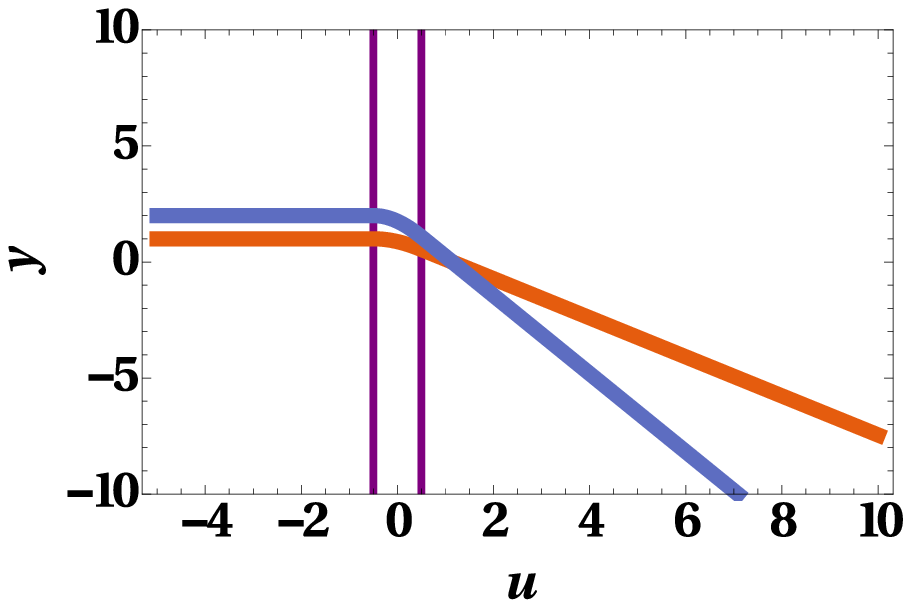}
   \caption{}
   \label{fig:y_plus}
  \end{subfigure}
  \caption{{Displacement memory effect along (a) $x$ and (b) $y$ directions for the first (orange, $x=1,y=1$) and second (blue, $x=2,y=2$) geodesics respectively, for a square pulse with values of $A_0=1,a=0.5$. The vertical lines in purple denote the sandwiched wave region in between two flat spacetimes (this is true for the next plot too).}}
 \label{fig:Disp_memory_plus}
\end{figure}

\noindent The memory effect can be easily realized in the above class of spacetimes by choosing suitable pulse profiles \cite{zhang1,zhang2}. A simple example is a square pulse with analytical form chosen as $A_+(u)=2A_0^2[\Theta(u+a)-\Theta(u-a)]$. The solutions are obtained by solving the geodesic equations and then matching them at the boundaries.\footnote{We have chosen $A_\times$ equal to zero.} Initially parallel geodesics before the wave region (purple vertical lines showing the boundary of the wave region) are seen to have a nonzero finite separation even after the passage of the pulse visible in the Figs. \ref{fig:x_plus} and \ref{fig:y_plus}.

\noindent Velocity memory effect is simply obtained by taking the first derivative of the solutions for geodesic equations. The results appear in Fig. 2, as shown.

\begin{figure}[H]
 \centering
  \begin{subfigure}[t]{0.45\textwidth}
   \centering
   \includegraphics[width=\textwidth]{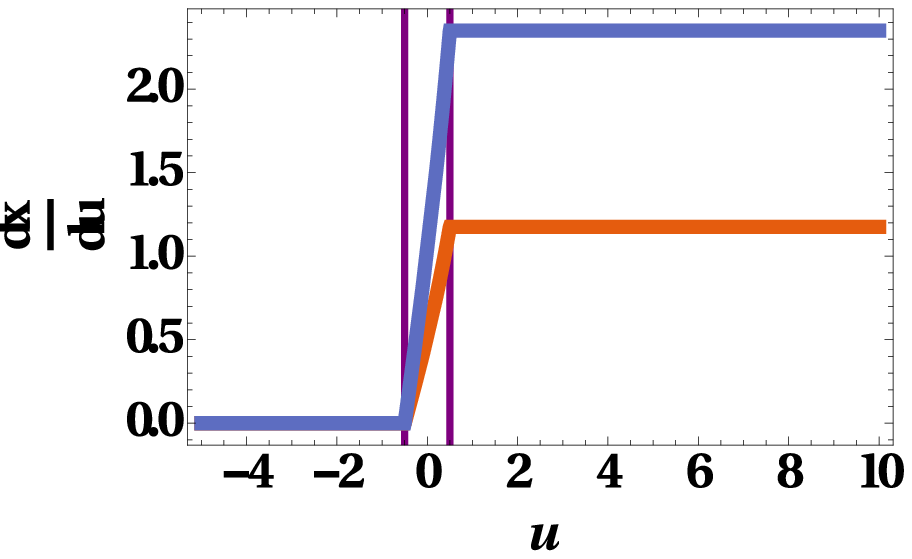}
   \caption{}
   \label{fig:x_vel_plus}
  \end{subfigure}\hspace{1.5cm}
  \begin{subfigure}[t]{0.45\textwidth}
   \centering
   \includegraphics[width=\textwidth]{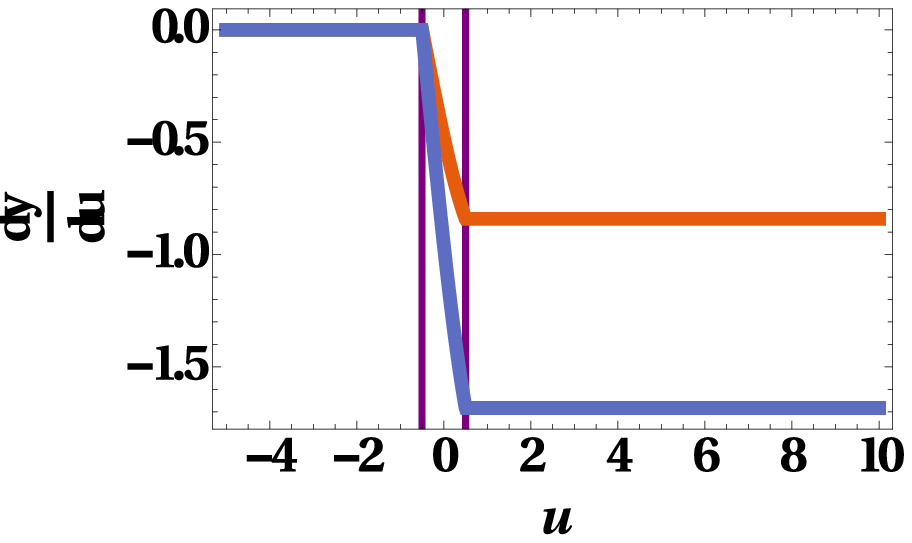}
   \caption{}
   \label{fig:y_vel_plus}
  \end{subfigure}
  \caption{{Velocity memory effect along (a) $x$ and (b) $y$ directions for the first (orange) and second (blue) geodesics  respectively. Here, a square pulse with $A_0=1$, $a=0.5$ is used.}}
 \label{fig:Vel_memory_plus}
\end{figure}

\noindent The velocity memory effect, as shown, displays a sharp change in the wave region which settles to a nonzero finite value. This is due to the fact that the pulse profile itself is discontinuous and hence $\ddot{x}(u)$ and $\ddot{y}(u)$ are discontinuous--a fact which follows from their geodesic equations (\ref{eq:combined_x}) and (\ref{eq:combined_y}).

\noindent In the following sections, we attempt to explore the possibility of arriving at a somewhat different memory effect from the evolution of the ${\cal B}$-tensor.

  

\section{Expansion, shear and rotation in Brinkmann coordinates for exact plane gravitational waves}
 
 \noindent The general formalism for obtaining the evolution of the kinematic variables in two dimensions (i.e., the Raychaudhuri equations) is available in \cite{r.sheikh}. The gradient of velocity [found by differentiation of $x(u)$ and $y(u)$ with respect to $u$] can be written as a second rank tensor which can be decomposed into expansion (trace), shear (symmetric, traceless) and rotation (antisymmetric):
\begin{align}
   {\cal B}_{ij}=\partial_jv_i=\begin{pmatrix} 
\frac{1}{2}\theta & 0 \\
0 & \frac{1}{2}\theta
\end{pmatrix}+\begin{pmatrix} 
\sigma_+ & \sigma_\times \\
\sigma_\times & -\sigma_+
\end{pmatrix}+\begin{pmatrix} 
0 & \omega \\
-\omega & 0\label{eq:twodimension}
\end{pmatrix}.
\end{align}
\noindent The evolution equation for the gradient of velocity may be written as
\begin{gather}
  v^k\partial_k(\partial_jv^i)=\partial_jf^i-(\partial_jv^k)(\partial_kv^i),\label{eq:velocityevolution1}\\  
  v^k\partial_k({\cal B}^i_{\,\,j})=\partial_jf^i-{\cal B}^i_{\,\,k} {\cal B}^k_{\,\,j}.\label{eq:velocityevolution2}
\end{gather}
In Eqs. (\ref{eq:velocityevolution1}) and (\ref{eq:velocityevolution2}) the term $f^i$ denotes the acceleration per unit mass. Hence $f^x=\frac{d^2x}{du^2}$ and $f^y=\frac{d^2y}{du^2}$. The four kinematic variables \{$\theta,\sigma_+,\sigma_\times,\omega$\} obey the evolution equation (\ref{eq:velocityevolution2}) which leads to separate
equations given as
\begin{gather}
    \frac{d\theta}{du}+\frac{\theta^2}{2}+2(\sigma_+^2+\sigma_\times^2-\omega^2)=\partial_xf^x+\partial_yf^y, \label{eq:thetaevolution} \\
    \frac{d\sigma_+}{du}+\theta\sigma_+=\frac{1}{2}(\partial_xf^x-\partial_yf^y), \label{eq:sigmaplusevolution} \\
    \frac{d\sigma_\times}{du}+\theta\sigma_\times=\frac{1}{2}(\partial_yf^x+\partial_xf^y), \label{eq:sigmacrossevolution} \\
    \frac{d\omega}{du}+\theta\omega=\frac{1}{2}(\partial_yf^x-\partial_xf^y). \label{eq:omegaevolution}
\end{gather}

\noindent Eqs. (\ref{eq:thetaevolution})-(\ref{eq:omegaevolution}) may be solved for specific pulse profiles (square pulse, sech-squared pulse and its derivatives) by substituting the values of $f^x$ and $f^y$ from their respective geodesic equations given in Eqs. (\ref{eq:combined_x}) and (\ref{eq:combined_y}).

\noindent The two pulse profiles that we choose to work with in this paper are as follows: \\
1. Square pulse, $A_+(u)=2A_0^2[\Theta(u+a)-\Theta(u-a)]$.\\
\noindent 2. Sech-squared pulse, $A_+(u)=\frac{1}{2}\sech^2(u)$.\\
The nature of the pulses vary in their differentiable nature. We further study consequences for the derivatives of the continuous sech-squared pulse. 
\subsection{Qualitative analysis on the evolution of kinematic variables}

\subsubsection{Generic pulse}\label{sssec:generic_analysis}
\noindent For any generic pulse having both polarizations $A_+(u)$ and $A_\times(u)$ be non-zero, the four evolution equations for these kinematic variables become
\begin{gather}
    \frac{d\theta}{du}+\frac{\theta^2}{2}+2(\sigma_+^2+\sigma_\times^2-\omega^2)=0, \label{eq:theta_evolution_gen} \\
    \frac{d\sigma_+}{du}+\theta\sigma_+=\frac{1}{2}A_+(u), \label{eq:sigma_plus_evolution_gen}\\
    \frac{d\sigma_\times}{du}+\theta\sigma_\times=\frac{1}{2}A_\times(u), \label{eq:sigma_cross_evolution_gen}\\
    \frac{d\omega}{du}+\theta\omega=0. \label{eq:omega_evolution_gen}
\end{gather}

\noindent A geodesic congruence starting out with initial values for all four variables set to zero simplifies the Eqs. (\ref{eq:theta_evolution_gen}-\ref{eq:omega_evolution_gen}). There is no twist in the entire range $u \hspace{2mm} \epsilon \hspace{2mm} (u_i, u_f)$ if $\omega=0$ initially. We consider only "+" polarization and therefore $\sigma_\times=0$ always. The resulting equations become
\begin{gather}
    \frac{d\theta}{du}+\frac{\theta^2}{2}+2\sigma_+^2=0, \label{eq:theta} \\
    \frac{d\sigma_+}{du}+\theta\sigma_+=\frac{1}{2}A_+(u). \label{eq:sigma_plus}
\end{gather}

\noindent Equation (\ref{eq:theta}) implies that $\frac{d\theta}{du}$ is always negative. Therefore, irrespective of its initial value at some $u_i$,
$\theta$ will eventually diverge to negative infinity. Equation (\ref{eq:sigma_plus}) is multiplied by two on both sides and added/subtracted to/from Eq. (\ref{eq:theta}) yielding a pair of uncoupled, first order ordinary differential equations. Then, defining $(\theta +2\sigma_+)=\xi$ and $(\theta-2\sigma_+)=\eta$, the corresponding equations turn out to be
 \begin{gather}
    \frac{d\xi}{du}+\frac{\xi^2}{2}=A_+(u), \label{eq:xi} \\
    \frac{d\eta}{du}+\frac{\eta^2}{2}=-A_+(u). \label{eq:eta}
\end{gather}

\noindent Note that the variables $\xi$ and $\eta$ are the
{\em eigenvalues} of the ${\cal B}$-matrix when $\sigma_\times$ and
$\omega$ are zero. From Eq. (\ref{eq:eta}) it is clear that $\dot\eta(u)<0$ as the function representing the pulse is manifestly positive and asymptotically zero in value. Integrating $\dot \eta(u)$ from $u_i$ 
($u_i$ will be negative and located far from the region around the origin where the pulse is nonzero in value) to $u_f$ (positive $u$ value and located far
down the positive $u$ axis), we conclude that since the right-hand side of Eq. (\ref{eq:eta}) is the negative of a positive definite number, the change \bigg[$\displaystyle{\lim_{u\to u_f}}\eta(u)-\displaystyle{\lim_{u\to u_i}} \eta(u)<0$\bigg]  is always less than zero. Mathematically, the change in the value of $\eta$ from $u\to u_i$ to $u\to u_f$ can never be set to zero due to the presence of the integral of the pulse (which equals the area enclosed between $u_i$
and $u_f$), and hence it is always nonzero, negative and finite.
But no such conclusion can be drawn for $\xi(u)$ from Eq. (\ref{eq:xi}). This equation can further be analyzed as yielding
\begin{align}
  \left. \xi \right|_{u_i}^{u_f} =-\frac{1}{2}\int_{u_i}^{u_f} \xi^2(u)\, du +\textrm{Area enclosed by the pulse}\, (=a_1).\label{eq:xi_1}
\end{align}

\noindent Equation (\ref{eq:xi_1}) implies that $\left. \xi \right|_{u_i}^{u_f}=0$ only if $\int_{u_i}^{u_f}\xi^2(u)\, du=2a_1$.

\noindent [In the following we will denote 
$\displaystyle{\lim_{u\to u_f}} P(u)$ (P may be $\theta, \sigma_+, \xi$ or  $\eta$) as simply $P(u_f)$, assuming its value to be zero when $u \to u_i$.] Thus, there are essentially two possibilities.

\begin{itemize}

    \item[] \textbf{(a)} $\xi(u_f)\geq0,\eta(u_f)<0$:
  
    \noindent The resulting inequalities are:
    
    $$\theta (u_f)-2\sigma_+(u_f)<0, \hspace{5mm} \theta (u_f)+2\sigma_+(u_f)\geq0.$$
    
    As is obvious from the above, $\theta \rightarrow -\infty$ and
    $\sigma_+ \rightarrow +\infty$ as $u\rightarrow u_f$ (focusing)
    is possible and will be shown as a consequence in
    various examples given later. On the other hand $\theta \rightarrow -\infty$
    and $\sigma_+ \rightarrow -\infty$ is not permissible because,
    as stated above, at $u=u_f$ (focusing), $\theta (u_f)+2\sigma_+(u_f) \geq 0$. 
    
    \item[] \textbf{(b)} $\xi(u_f)\leq0,\eta(u_f)<0$:

    \noindent The resulting inequalities are
    
    $$\theta (u_f)-2\sigma_+ (u_f)<0, \hspace{5mm} \theta (u_f)+2\sigma_+ (u_f)\leq0.$$
    Hence,
    \begin{equation}
          \theta(u_f)<0, \hspace{5mm} \Bigg(\sigma_+ (u_f)>\frac{\theta (u_f)}{2}\Bigg) \hspace{1.5mm} \textrm{or}  \hspace{2mm} \Bigg(\sigma_+(u_f)\leq-\frac{\theta(u_f)}{2}\Bigg).   \label{eq:case_b}
    \end{equation}
    Thus, $\theta \rightarrow -\infty$, $\sigma_+ \rightarrow +\infty$
    as well as $\theta \rightarrow -\infty$, $\sigma_+ \rightarrow -\infty$ are both permissible, modulo constraints. We will see
    examples of this too, later. 
    
\end{itemize}

\subsubsection{Derivatives of the pulse}

\noindent Let us now analyze the nature of evolution of the kinematical variables for the first three derivatives of the pulse. The physical relevance which motivates us to look at consequences for the derivatives of a pulse are given in the following section (Sec. \ref{sec:derivatives}). Equations (\ref{eq:theta}) and (\ref{eq:sigma_plus}) are now modified on their rhs's with the pulse being replaced by its derivative
(first, second or third). Since we consider an even pulse (sech-squared pulse), the first and third derivatives are odd functions and the second derivative is even.

\begin{itemize}
    \item \textbf{First derivative and third derivative:}
    
    \noindent Let us consider the case for the first derivative. The new set of equations are
    \begin{gather}
    \frac{d\xi}{du}+\frac{\xi^2}{2}=\dfrac{dA_+}{du}, \label{eq:xi_1stdv} \\
    \frac{d\eta}{du}+\frac{\eta^2}{2}=-\dfrac{dA_+}{du}. \label{eq:eta_1stdv}
    \end{gather}
    
    \noindent We employ the same trick as was done in the case of the pulse.
    The integral of the first derivative of the pulse is positive as long as we assume 
    $u_i$ far down the negative $u$ axis and $u_f$ relatively closer to $u=0$ on the
    positive $u$ direction, but reasonably away from the region where the function is clearly nonzero. Thus, in this case we end up with conclusions similar to the case of the pulse discussed just above.
    
    The same line of argument holds for the third derivative too (odd function).
     
    \item \textbf{Second derivative:} 

 \noindent In this case, the resulting equations are:    
    \begin{gather}
    \frac{d\xi}{du}+\frac{\xi^2}{2}=\dfrac{d^2A_+}{du^2}, \label{eq:xi_2nddv} \\
    \frac{d\eta}{du}+\frac{\eta^2}{2}=-\dfrac{d^2A_+}{du^2}. \label{eq:eta_2nddv}
    \end{gather}
    
    \noindent The second derivative of the pulse is an even function. 
    The integral of the second derivative from $u_i$ to $u_f$ can be 
    seen to be negative for a sech-squared pulse though other examples
    do exist. We discuss this somewhat opposite behaviour because it is
    different. Here, we will find that, at $u_f$, $\xi (u_f) < 0$ but
    $\eta (u_f)$ may be greater or less than zero. Thus, we have
    
    $$\theta (u_f) + 2\sigma_+ (u_f) <0, \hspace{5mm} \theta (u_f) -2 \sigma_+ (u_f) >0.$$
    
    Hence, it is clear that $\theta \rightarrow -\infty$, $\sigma_+ \rightarrow -\infty$ is allowed, though $\theta \rightarrow -\infty$, $\sigma_+ \rightarrow +\infty$ is not.  On the contrary, if $\xi<0$, $\eta<0$, then
    
    $$\theta (u_f) + 2\sigma_+ (u_f) <0, \hspace{5mm} \theta (u_f) -2 \sigma_+ (u_f)<0.$$

    Here, $\theta \rightarrow -\infty$ and $\sigma_+ \rightarrow -\infty$  and $\theta \rightarrow -\infty$ and $\sigma_+ \rightarrow +\infty$ may both arise and are allowed.

\end{itemize}

\noindent We will illustrate some of the above-mentioned features 
related to the case of the second derivative of a pulse,
when we discuss the  example of a sech-squared pulse later.

\noindent We now move on to specific examples in the following section.

\subsection{Kinematic variables for square pulse in plus polarization}
\noindent The results given below for the square pulse are fully analytical. The values of $f^x$ and $f^y$ are obtained from the geodesic equations of the pulse profile. We have
\begin{align}
f^x=\frac{d^2x}{du^2}=\begin{cases} 
  0 & u\leq -a\\ 
  A_0^2x & -a\leq u\leq a,\\
  0 & u\geq a
  \end{cases} \label{eq:fx}
\end{align}

 \begin{align}
  f^y=\frac{d^2y}{du^2}=\begin{cases} 
  0 & u\leq -a\\ 
  -A_0^2y & -a\leq u\leq a.\\
  0 & u\geq a
  \end{cases} \label{eq:fy}
 \end{align}
 Equations (\ref{eq:thetaevolution})-(\ref{eq:omegaevolution}) in the first ($u\leq-a$) and third ($u\geq a$) regions have zero value on their rhs's. We assume initial values of all the kinematic variables to be zero. So, $\sigma_\times$ and $\omega$ become zero in all the regions. Thus, the evolution equations in the second region ($-a<u<a$) become
 \begin{gather}
    \frac{d\theta}{du}+\frac{\theta^2}{2}+2\sigma_+^2=0, \label{eq:thetasq} \\
    \frac{d\sigma_+}{du}+\theta\sigma_+=A_0^2. \label{eq:sigmaplussq} 
\end{gather}

\noindent Hence, using the transformed variables $\xi$ and $\eta$ we solve Eqs. (\ref{eq:xi}) and (\ref{eq:eta}) in all three regions. The solutions in  the second (wave) region turn out as
\begin{align}
    \xi=2A_0\tanh[A_0(u+C_1)],\label{eq:xieqn}\\
  \eta=2A_0\tan[A_0(C_2-u)].\label{eq:etaeqn}
\end{align}
\noindent Thus, $\theta$ and $\sigma_+$  can found from $\xi$ and $\eta$. Thereafter, matching values at $u=-a$ (which is zero for both variables) the value for $C_1,C_2$ is obtained. In the
same way, the solution for the region $u\geq a$ is obtained by matching at $u=a$. The final solutions for expansion and shear are
\begin{align}
  \theta(u)=\begin{cases} 
  0 & u\leq -a\\ 
  A_0[\tanh[A_0(u+a)]-\tan[A_0(u+a)]] & -a\leq u\leq a,\\
  \frac{1}{u-a+A_0^{-1}\coth[2aA_0]}+\frac{1}{u-a-A_0^{-1}\cot[2aA_0]} & u\geq a
  \end{cases} \label{eq:thetasolnsqp}
\end{align}
 \begin{align}
  \sigma_+(u)=\begin{cases}
   0 & u\leq -a \\
    \frac{A_0}{2}[\tanh[A_0(u+a)]+\tan[A_0(u+a)]] & -a\leq u \leq a.\\
   \frac{1}{2}\Big(\frac{1}{u-a+A_0^{-1}\coth[2aA_0]}-\frac{1}{u-a-A_0^{-1}\cot[2aA_0]}\Big) & u\geq a
  \end{cases} \label{eq:sigmaplussoln}
 \end{align}
 \begin{figure}[H]
	\centering
	\begin{subfigure}[t]{0.45\textwidth}
		\centering
		\includegraphics[width=\textwidth]{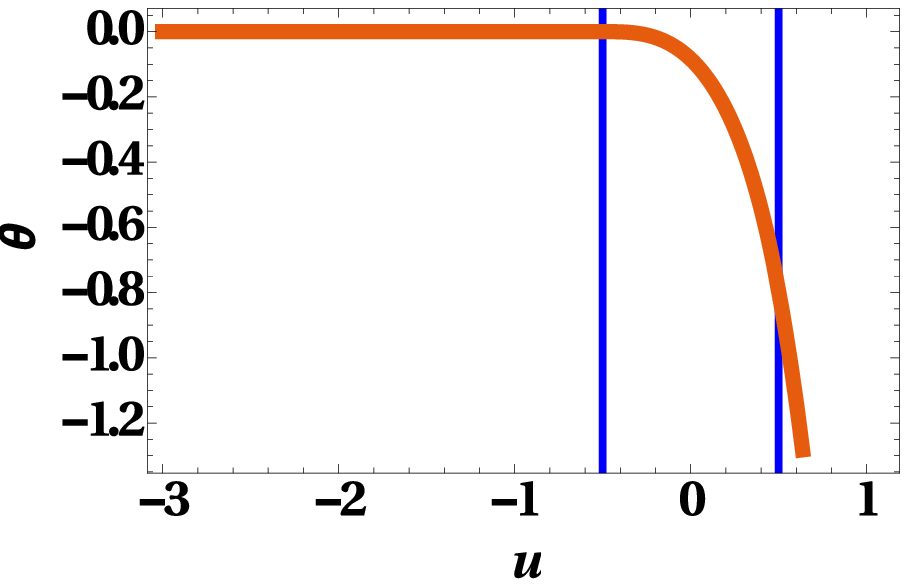}
		 \caption{}
		\label{fig:exp_sq_plus_}
\end{subfigure}\hspace{1cm}
	\begin{subfigure}[t]{0.45\textwidth}
		\centering
		\includegraphics[width=\textwidth]{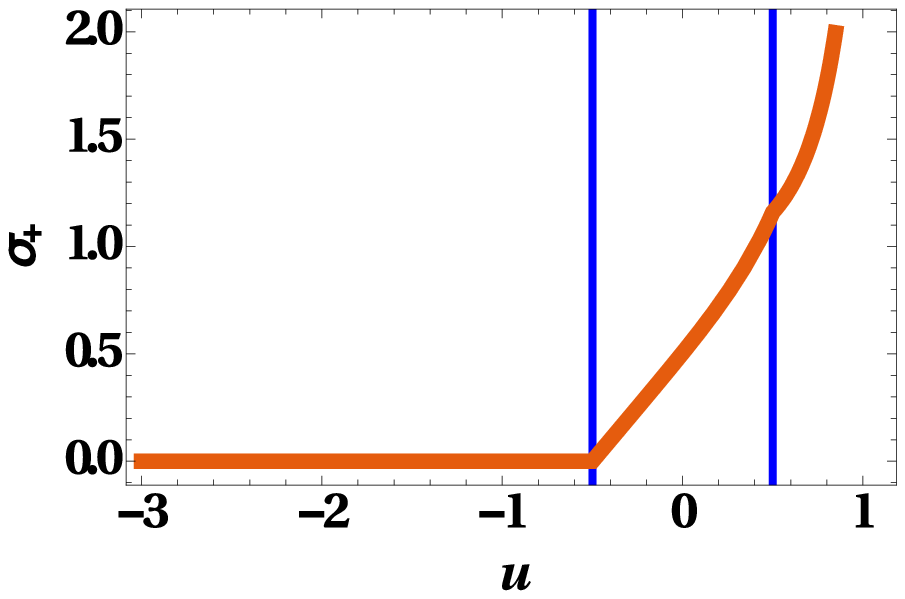}
		 \caption{}
		\label{fig:shear_sq_plus}
	\end{subfigure}
	\caption{(a) Expansion and (b) shear variation in case of square pulse for $A_0=1,a=0.5$. The plot for shear shows piecewise smoothness owing to Eq. (\ref{eq:sigma_plus_evolution_gen}). The vertical lines in blue demarcates the wave region from the flat spacetime region.}
	\label{fig:squarepulseplots}
\end{figure}
\noindent The plots in  Figs. \ref{fig:exp_sq_plus_} and \ref{fig:shear_sq_plus} have kinks at $u=-0.5,0.5$  because of the nature of $A_+(u)$. The expansion $\theta$ develops a
negativity as $u$ enters the region where the pulse is nonzero. This acquired negativity drives it
toward a focal point after the pulse has departed, i.e., beyond $u=a$. The appearance of 
the focal point is what we noted earlier as the intersection of the
geodesics beyond $u=a$.  Similarly, the initially zero shear acquires a positivity on entering the region where the pulse is active and nonzero. Subsequently, even after the departure of the pulse the shear keeps on increasing. Thus, there is a permanent change in the shear and expansion (after the pulse departs), which is in contrast to their zero value before the arrival of the pulse. Further, we note that $\theta
\to-\inf$ at $u=a+A_0^{-1}\cot[2aA_0]$, which {\em clearly depends on the width and height of the pulse}. 

\noindent It turns out that $\xi\geq0$ and $\eta<0$ in region 3 (i.e., $u\geq a$), thereby obeying the inequality obtained for case (a) in (Sec. \ref{sssec:generic_analysis}). 
\noindent The solutions and conclusions for cross polarization are exactly the same as for plus polarization with $\sigma_+$ replaced by $\sigma_\times$.

\subsection{Kinematic variables for sech-squared pulse in plus and cross polarizations}

\noindent Here, the pulse profile is a continuous function and hence the set of two coupled equations can be solved for the entire range of the affine parameter. We show here the results for plus polarization (for $\times$ polarization the analytical results and plots are similar). From
the geodesic equations, we obtain $f^x=\frac{1}{4}x\sech^2(u)$ and $f^y=-\frac{1}{4}y\sech^2(u)$ for the pulse profile. Subsequently, we solve Eqs. (\ref{eq:theta}) and (\ref{eq:sigma_plus}) by using Eqs.(\ref{eq:xi}) and (\ref{eq:eta}): \footnote{The same relationships between the variables $\{\theta,\sigma_+\}$ and $\{\xi, \eta\}$ are used.}
\begin{gather}
    \frac{d\xi}{du}+\frac{1}{2}{\xi^2}=\frac{1}{2}\sech^2(u), \label{eq:newthetasech} \\
    \frac{d\eta}{du}+\frac{1}{2}{\eta^2}=-\frac{1}{2}\sech^2(u). \label{eq:newsigmaplussech} 
\end{gather}

\noindent Equations (\ref{eq:newthetasech}) and (\ref{eq:newsigmaplussech}) can be solved analytically by the substitution, $\xi=2\dot{\alpha}/\alpha, \eta=2\dot{\beta}/\beta$, which leads to:
\begin{gather}
    \frac{d^2\alpha}{du^2}=\frac{1}{4}\sech^2(u) \alpha, \label{eq:substitution_sech1} \\
    \frac{d^2\beta}{du^2}=-\frac{1}{4}\sech^2(u) \beta. \label{eq:substitution_sech2} 
\end{gather}

\noindent The solutions of Eqs. (\ref{eq:substitution_sech1}) and (\ref{eq:substitution_sech2}) are
\begin{gather}
    \alpha(u)=C_1 \,K\bigg[\dfrac{1}{2}\big(1-\tanh(u)\big)\bigg]+C_2 \,Q_{-\frac{1}{2}}\big(\tanh(u)\big), \label{eq:substitution_sech3}\\
    \beta(u)=C_3 \,P_{\frac{1}{2}(\sqrt{2}-1)}\big(\tanh(u)\big)+ C_4 \,Q_{\frac{1}{2}(\sqrt{2}-1)}\big(\tanh(u)\big). \label{eq:substitution_sech4}
\end{gather}
 
\noindent where, in Eq. (\ref{eq:substitution_sech3}), the first function is a complete elliptic integral of the first kind and the second is a Legendre function of the second kind. In Eq. (\ref{eq:substitution_sech4}) we have Legendre functions of first and second kinds respectively. The relationship between the kinematic variables $\{\theta,\sigma_+\}$ and $\{ \alpha, \beta\}$ is
\begin{equation}
    \theta=\bigg(\frac{\dot{\alpha}}{\alpha}+\frac{\dot{\beta}}{\beta}\bigg),
    \hspace{2.5cm} \sigma_+=\dfrac{1}{2}\bigg(\frac{\dot{\alpha}}{\alpha}-\frac{\dot{\beta}}{\beta}\bigg).
    \label{eq:relation}
\end{equation}

\noindent Substituting back the functional forms of $\alpha,\beta$ as obtained from Eqs. (\ref{eq:substitution_sech3}) and (\ref{eq:substitution_sech4}) into Eq. (\ref{eq:relation}) we get analytic expressions of $\theta$ and $\sigma_+$. Since Eqs. (\ref{eq:substitution_sech1}) and (\ref{eq:substitution_sech2}) are second order ordinary differential equations, we end up with a total of four constants. However, our initial equations (\ref{eq:newthetasech}) and (\ref{eq:newsigmaplussech}) were first order and hence we should have two arbitrary constants. Hence, these are fixed by setting the value of $\theta, \sigma_+ $ to zero at an initial value of $u$ (i.e., we have an initially parallel geodesic congruence). Thus, among the four constants in Eqs. (\ref{eq:substitution_sech3}) and (\ref{eq:substitution_sech4}) we choose two freely and the other two are fixed by constraining $\dot{\alpha}=\dot{\beta}=0$ at an initial $u$.
The plots generated for the kinematic variables are shown in Fig. 4.

\begin{figure}[H]
	\centering
	\begin{subfigure}[t]{0.45\textwidth}
		\centering
		\includegraphics[width=\textwidth]{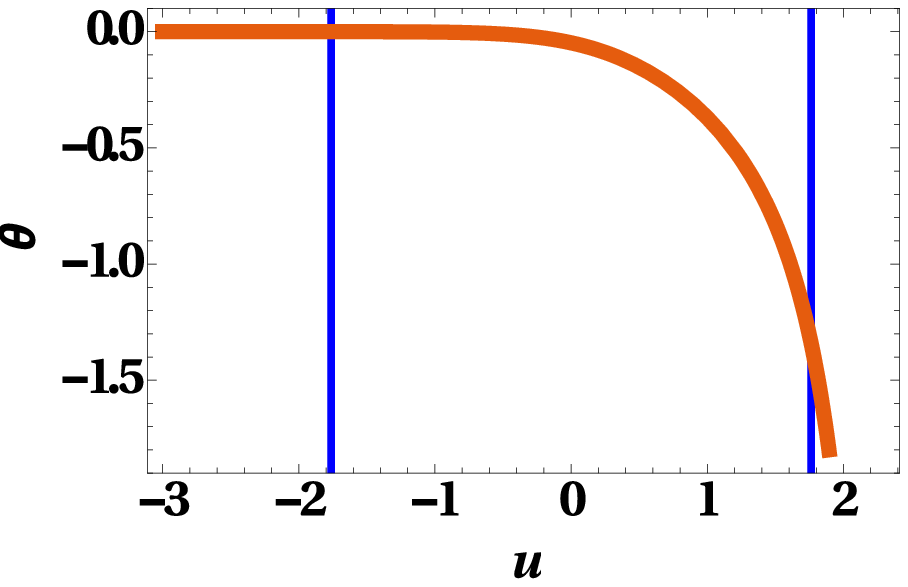}
		 \caption{}
		\label{fig:exp_sech_plus_analytic}
\end{subfigure}\hspace{1cm}
	\begin{subfigure}[t]{0.45\textwidth}
		\centering
		\includegraphics[width=\textwidth]{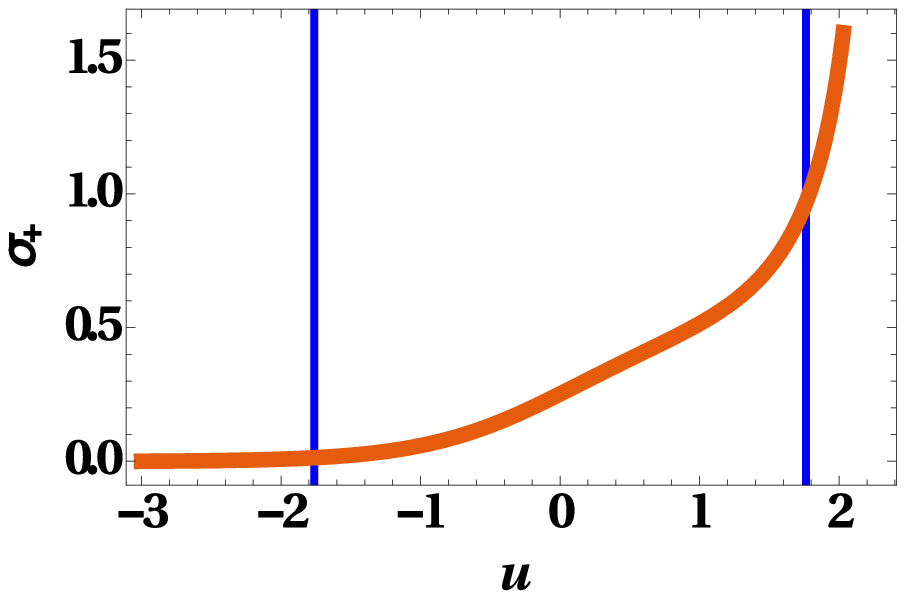}
		 \caption{}
		\label{fig:shear_sech_plus_analytic}
	\end{subfigure}
	\caption{(a) Expansion and (b) shear variation in case of sech-squared pulse for $C_1=C_3=1$. The blue vertical lines show the FWHM region of the pulse. The plots are continuous indicating the smooth nature of the pulse profile.}
	\label{fig:sechpulseplots_analytic}
\end{figure}

\noindent Conclusions from Figs. (\ref{fig:exp_sech_plus_analytic}) and (\ref{fig:shear_sech_plus_analytic}) are largely the same as found earlier for a square pulse. A permanent change in the expansion and shear of the congruence is noted here too. The plot for shear does not exhibit any kink (as seen in the square pulse)-- the smoothness
due to the continuous nature of the pulse.

\noindent In both cases (square and sech squared), expansion is always negative and $\sigma_+$ is monotonically increasing. It may
be checked that the constraints imposed from the analysis of a generic pulse as discussed in Sec. IIIA, hold.

\section{Derivatives of sech-squared pulse} \label{sec:derivatives}

\noindent We now move on toward applying the above formalism for calculating kinematic variables, when we have various derivatives of a sech-squared pulse. The nature of derivatives and its integrals over the duration of the pulse have been discussed previously in \cite{Gibbons} and much later by Zhang {\em et al}. in \cite{zhang}. In linear theory, the source quadrupole moment is related to the curvature tensor via the formula
 \begin{equation}
     R_{i0j0}=\frac{G}{3r}\dfrac{d^4D_{ij}}{dt^4}. \label{eq:quadrupole}
 \end{equation}

 \noindent The above-mentioned authors defined integrals over the Riemann tensor in the limit where the wave is localized and subsequently looked at their values for the first three derivatives of chosen pulse profiles: 
 
\begin{gather}
I^{(3)}=\int_{t_i}^{t_f}dt \int_{t_i}^{t}dt^{\prime} \int_{t_i}^{t^\prime} dt\mydprime\, R_{0i0j}(t\mydprime), \label{eq:1st_intergral}\\
I^{(2)}=\int_{t_i}^{t_f}dt \int_{t_i}^{t}dt^{\prime}\, R_{0i0j}(t\myprime), \label{eq:2nd_integral}\\
I^{(1)}=\int_{t_i}^{t_f}dt\, R_{0i0j}(t). \label{eq:3rd_integral}
\end{gather}

 \noindent Depending upon the physical scenario (such as, collapse or flybys) the initial and final quadrupole moment would differ and hence one can obtain the nature of an incoming pulse by analyzing the number of times the Riemann tensor has changed sign (i.e., by evaluating $I^{(1)}$, $I^{(2)}$ and $I^{(3)}$). Even in full nonlinear theory, one can guess the approximate nature of a pulse by knowing the values of these integrals, although Eq.(\ref{eq:quadrupole}) does not hold. Here we calculate these integrals for the derivatives of a sech-squared pulse and explain the corresponding physical scenario. 
 Subsequently, we analyze the nature of expansion and shear for geodesic congruences, in the presence of the various derivatives of the
 sech-squared pulse.

\begin{itemize}[leftmargin=*]
    
 \item[] \textbf{(a) First derivative:} 
 
  \noindent Recall the pulse profile given before. Its derivative would
  lead to (Fig. 5),
      \begin{equation}
                A_1(u)= \frac{dA_+(u)}{du}=\frac{1}{2}\dfrac{d}{du}\sech^2(u)=-\sech^2(u)\tanh(u).\label{eq:1st_derivative}
      \end{equation}    
      
      \begin{figure}[H]
          \centering
          \includegraphics[scale=0.9]{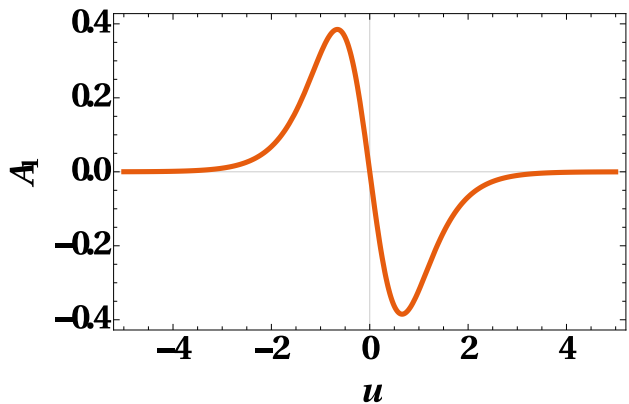}
          \caption{First derivative of sech-squared pulse. This corresponds to a flyby scenario where the gravitational radiation is emitted via bremsstrahlung.}
          \label{fig:1st_derivative}
      \end{figure}
\noindent The integrals given by Eqs. (\ref{eq:1st_intergral})-(\ref{eq:3rd_integral}) are evaluated for this pulse. The values are
 \begin{equation}
 I^{(1)}=0, \hspace{2cm} I^{(2)}=1, \hspace{2cm} I^{(3)} \to \infty.
 \end{equation}
 This could correspond to the case of a flyby leading to gravitational bremsstrahlung. Kovacs and Thorne \cite{kovacs} gave an analytic expression for the metric perturbation as a function of time and other parameters related to the binary (viz., mass, inclination angle, impact parameter). At initial times ($t \to -\infty$), the dominant contribution is constant. Hence, the quadrupole moment at initial instant is proportional to a quadratic function of time. Thus, both $I^{(2)}$ and $I^{(3)}$ are nonzero, following from Eq. (\ref{eq:quadrupole}). The non-zero kinematic variables are solved numerically from Eqs.(\ref{eq:theta}) and (\ref{eq:sigma_plus})\footnote{\label{note}$A_+(u)$ is now replaced with $A_1(u)$. Also, in the case of second and third derivatives $A_+(u)$ is replaced by $A_2(u)$ and $A_3(u)$ respectively on the right-hand side of Eq. (\ref{eq:sigma_plus}).} in {\em Mathematica 10} and are shown in the plots of Fig. 6
 \begin{figure}[H]
	\centering
	\begin{subfigure}[t]{0.45\textwidth}
		\centering
		\includegraphics[width=\textwidth]{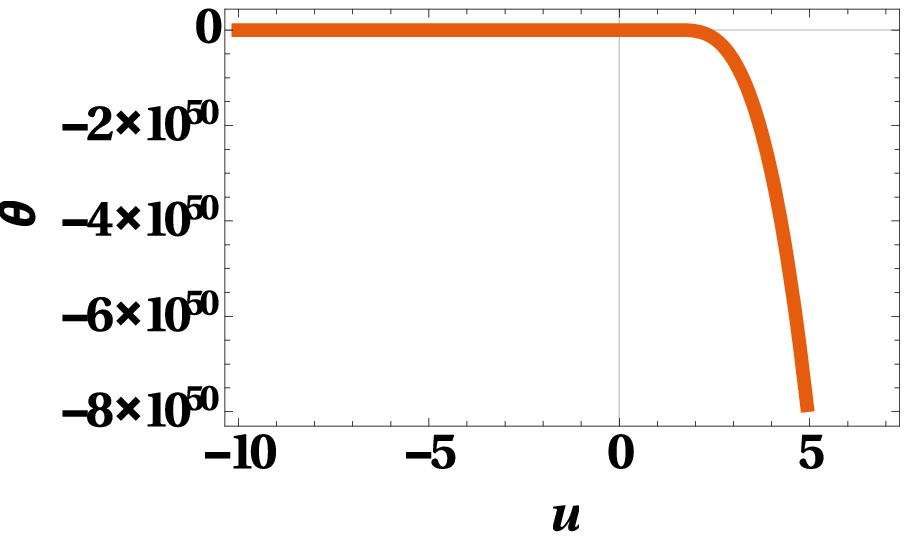}
		 \caption{}
		\label{fig:exp_1st_dv_sech_plus}
\end{subfigure}\hspace{1cm}
	\begin{subfigure}[t]{0.45\textwidth}
		\centering
		\includegraphics[width=\textwidth]{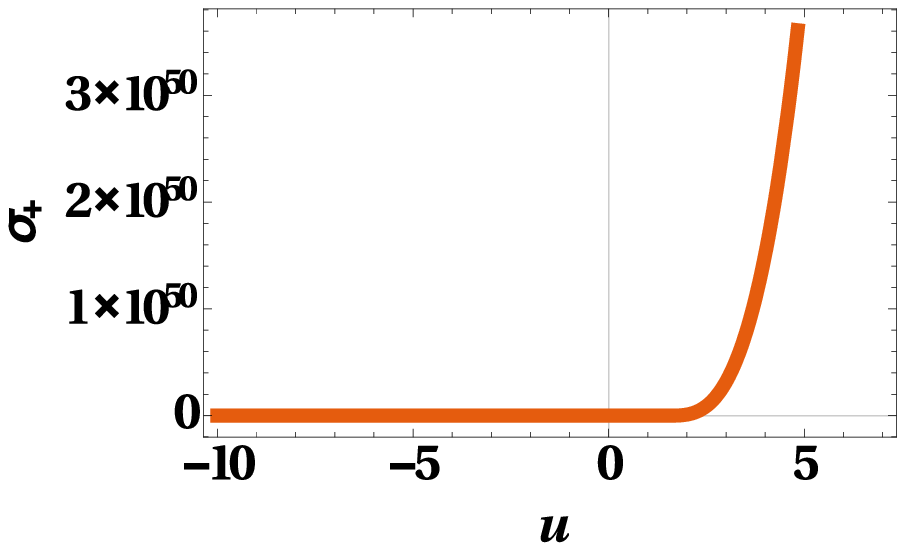}
		 \caption{}
		\label{fig:shear_1st_dv_sech_plus}
	\end{subfigure}
	\caption{(a) Expansion ($-10<u<7$) and (b) shear ($-10<u<7$) variation in case of a first derivative of a sech-squared pulse. The initial value of both expansion and shear is set to zero (This is also true for second and third derivatives).}
	\label{fig:1st_detivative_kv}
\end{figure}

 \noindent Figs.(\ref{fig:exp_1st_dv_sech_plus}) and (\ref{fig:shear_1st_dv_sech_plus}) are in accordance with the constraint imposed by the condition $\theta \rightarrow-\infty$ and $\sigma_+\rightarrow +\infty$ as discussed previously in Sec. III A.

\item[] \textbf{(b) Second derivative:} 
 
  \noindent In this case, the pulse profile (see Fig. 7) is               
      \begin{equation}
             A_2(u)=   \frac{d^2A_+(u)}{du^2}=\frac{1}{2}\dfrac{d^2}{du^2}\sech^2(u)=-\sech^4(u)+2\tanh^2(u)\sech^2(u).\label{eq:2nd_derivative}
      \end{equation}  
      
      \begin{figure}[H]
          \centering
          \includegraphics[scale=0.8]{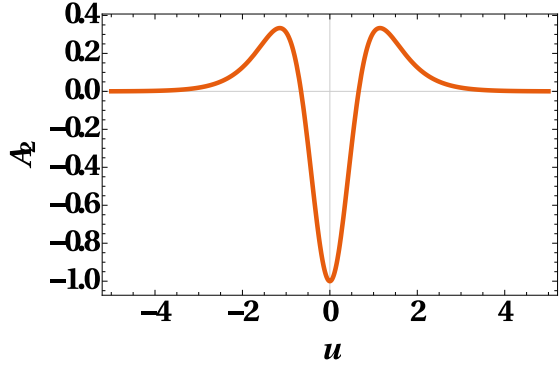}
          \caption{Second derivative of a sech-squared pulse.}
          \label{fig:2nd_derivative}
      \end{figure}
      
\noindent As done for the first derivative, we find the integrals given by Eqs. (\ref{eq:1st_intergral})-(\ref{eq:3rd_integral}) for the pulse given in Eq. (\ref{eq:2nd_derivative}). The values are
 \begin{equation}
 I^{(1)}=0, \hspace{2cm} I^{(2)}=0, \hspace{2cm} I^{(3)}=1.
 \end{equation}
 
 \noindent This scenario was considered by Braginsky and Thorne \cite{nature} where they distinguished between bursts with and without memory within linearized gravity. In the latter case, the metric perturbation vanishes beyond the wave region and hence $I^{(2)}$ vanishes and thus no memory effect is possible\footnote{Exact plane wave spacetimes are exact solutions in full nonlinear general relativity. Hence, even in this case we observe a memory effect.}. In order to have a finite value of $h_{ij}$ beyond the wave region, $I^{(2)}$ has to be nonzero and finite.
\noindent The plots are obtained after solving numerically\footref{note} in {\em Mathematica 10}.

 \begin{figure}[H]
	\centering
	\begin{subfigure}[t]{0.4\textwidth}
		\centering
		\includegraphics[width=\textwidth]{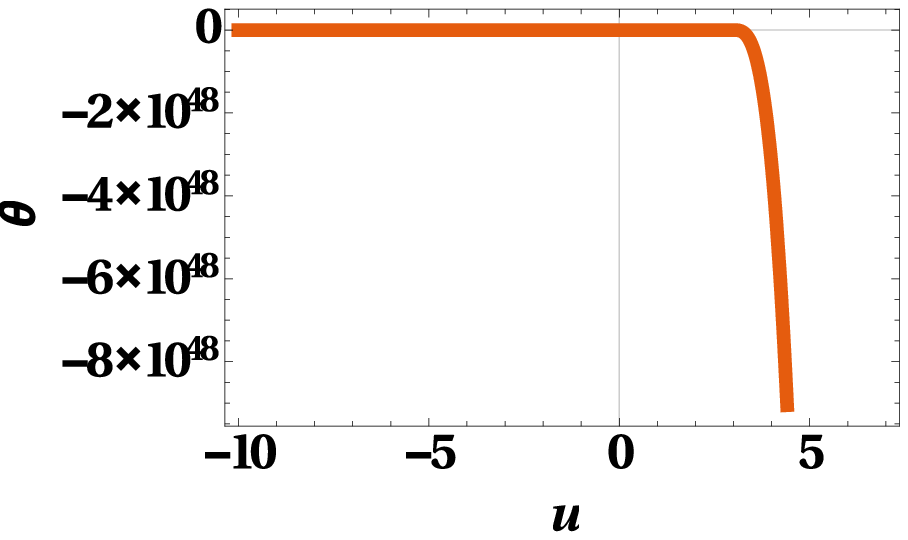}
		 \caption{}
		\label{fig:exp_2nd_dv_sech_plus}
\end{subfigure}\hspace{1cm}
	\begin{subfigure}[t]{0.4\textwidth}
		\centering
		\includegraphics[width=\textwidth]{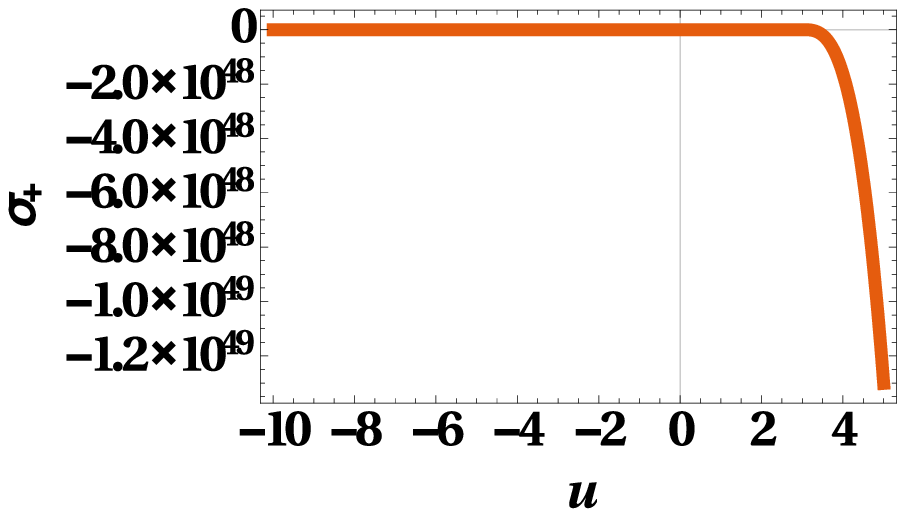}
		 \caption{}
		\label{fig:shear_2nd_dv_sech_plus}
	\end{subfigure}
	\caption{(a) Expansion ($-10<u<7$) and (b) shear ($-10<u<7$) variation in case of a second derivative of a sech-squared pulse.}
	\label{fig:2nd_detivative_kv}
\end{figure}
\noindent In this case, both expansion and shear [see Figs. 8a and 8b] diverge to minus infinity,
which is permissible from the generic analysis (Sec. III A).

\item[] \textbf{(c) Third derivative:} 
 
  \noindent In this case, the pulse profile (Fig. 9) is given as:               
      \begin{equation}
                A_3(u)= \dfrac{d^3A_+(u)}{du^3}=\frac{1}{2}\dfrac{d^3}{du^3}\sech^2(u)=8\sech^4(u)\tanh(u)-4\tanh^3(u)\sech^2(u).\label{eq:3rd_derivative}
      \end{equation}    
      
      \begin{figure}[H]
          \centering
          \includegraphics[scale=0.6]{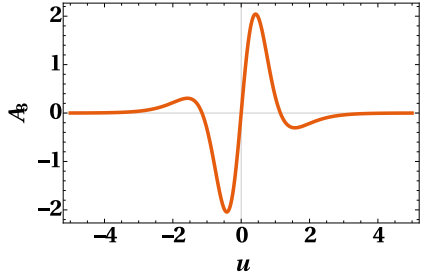}
          \caption{Third derivative of sech-squared pulse. This correspond to the case of gravitational collapse.}
          \label{fig:3rd_derivative}
      \end{figure}     
\noindent The integrals for this pulse [as given in Eq. (\ref{eq:3rd_derivative})] become
 \begin{equation}
 I^{(1)}=0, \hspace{2cm} I^{(2)}=0, \hspace{2cm} I^{(3)}=0.
 \end{equation}
 This scenario is of gravitational collapse \cite{Gibbons}. The quadrupole moment tensor is initially and finally time independent. Hence, the  first derivative of the quadrupole moment tensor vanishes. From Eq. (\ref{eq:quadrupole}) one finds that $I^{(3)}$ vanishes. This implies that the  minimum number of turning points for a pulse from gravitational collapse has to have at least three sign changes.
 The numerical solutions obtained after solving Eqs. (\ref{eq:theta}) and (\ref{eq:sigma_plus})\footref{note} leads to the plots in Fig. 10.
 \begin{figure}[H]
	\centering
	\begin{subfigure}[t]{0.3\textwidth}
		\centering
		\includegraphics[width=\textwidth]{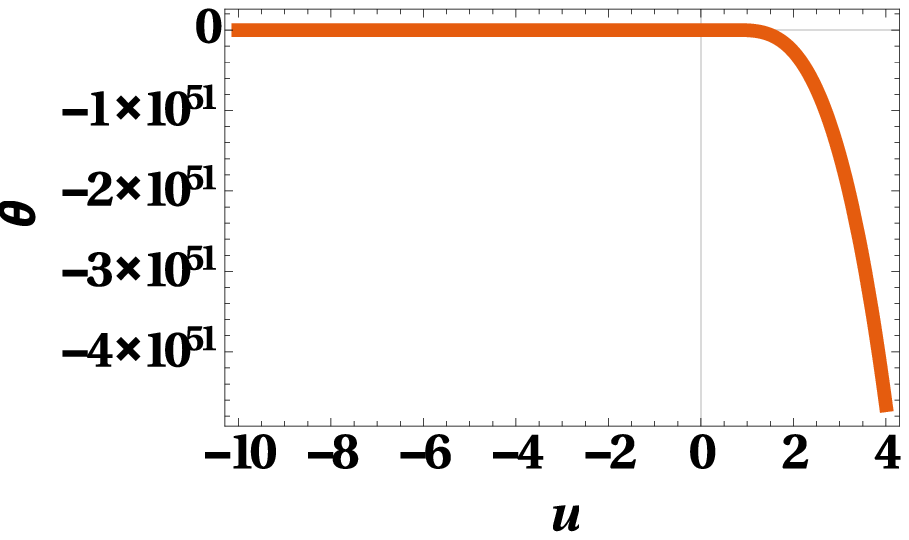}
		 \caption{}
		\label{fig:exp_3rd_dv_sech_plus}
\end{subfigure}\hspace{1.5cm}
	\begin{subfigure}[t]{0.3\textwidth}
		\centering
		\includegraphics[width=\textwidth]{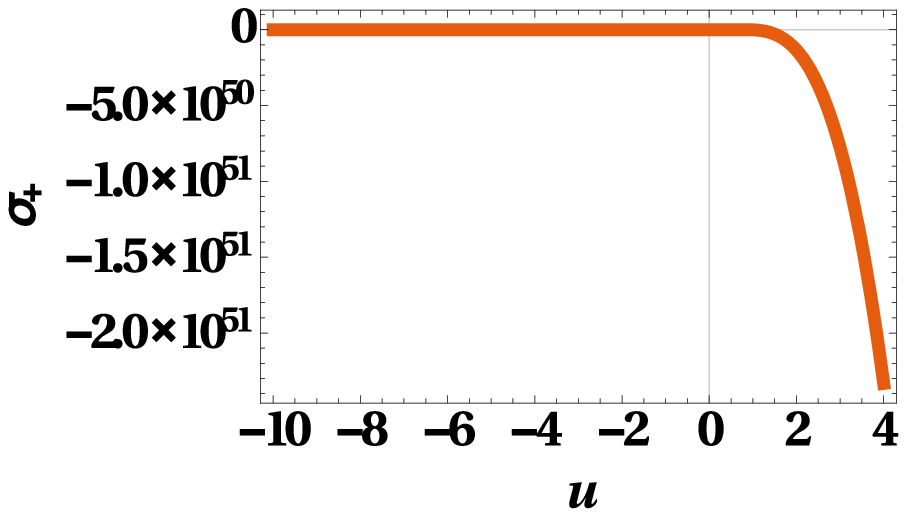}
		 \caption{}
		\label{fig:shear_3rd_dv_sech_plus}
	\end{subfigure}
	\caption{(a) Expansion ($-10<u<4$) and (b) shear ($-10<u<4$) variation in case of the third derivative of a sech-squared pulse.}
	\label{fig:3rd_detivative_kv}
\end{figure}

\noindent The nature of the plots in Fig.(\ref{fig:3rd_detivative_kv}) also obey the constraint $\theta \rightarrow-\infty$ and $\sigma_+\rightarrow -\infty$. Here too, like in the case of the second derivative, both expansion and shear diverges to minus infinity (allowed via the
qualitative analysis in Sec. III A above).
\end{itemize}

\noindent Thus, expansion is relatively of the same nature in all these three cases (following from qualitative arguments) while no such definite constraint can be imposed on the sign of shear.

\noindent Plane gravitational waves form caustics and hence can act as gravitational lenses. This feature was initially studied by Penrose \cite{penrose} for null geodesics and has been renewed by Harte and Drivas \cite{Harte} for  better understanding of gravitational lensing from a theoretical point of view.

\begin{figure}[H]
	\centering
	\begin{subfigure}[t]{0.4\textwidth}
		\centering
		\includegraphics[width=\textwidth]{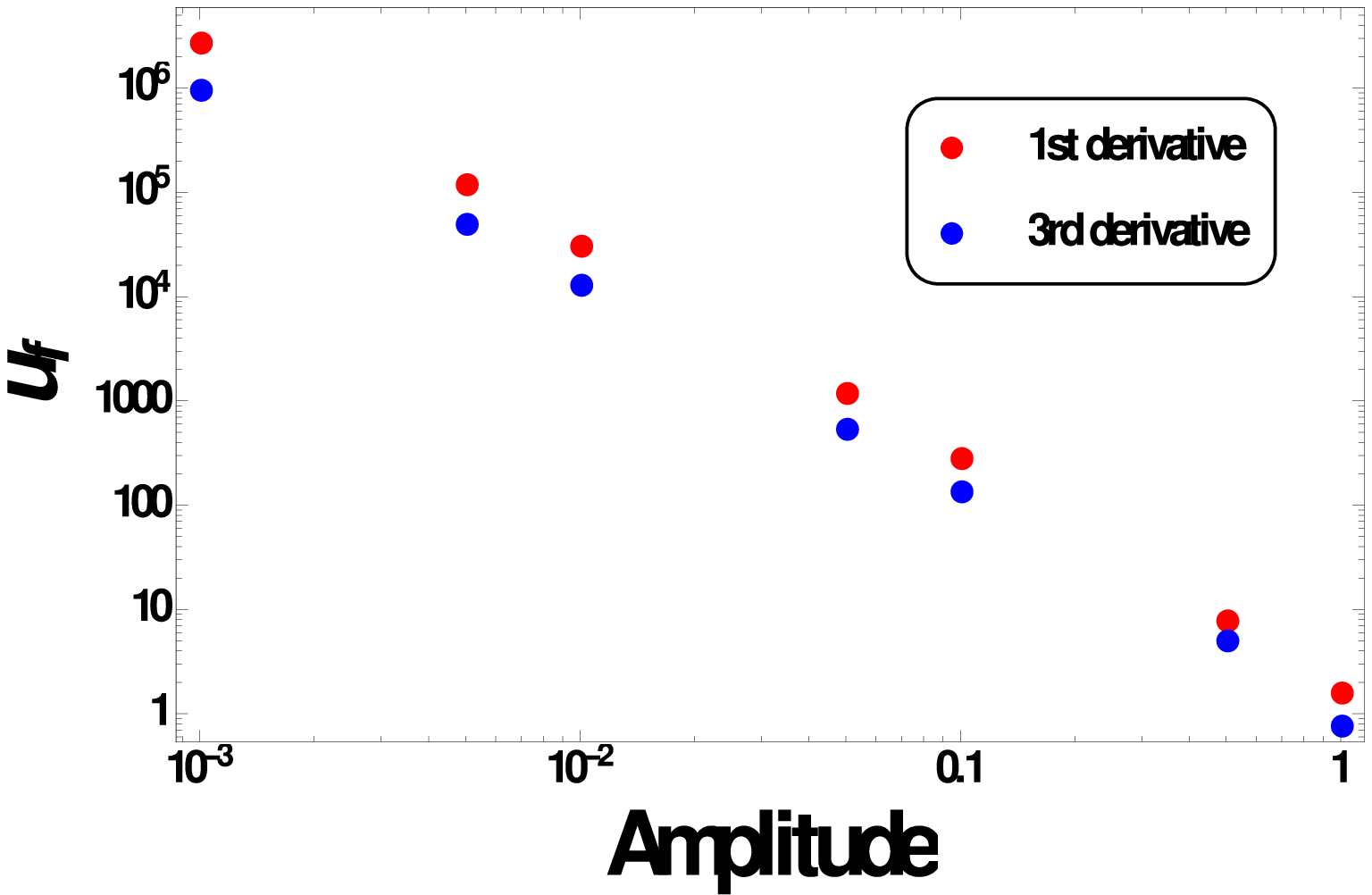}
		 \caption{}
		\label{fig:a_vs_uf}
\end{subfigure}\hspace{1cm}
	\begin{subfigure}[t]{0.4\textwidth}
		\centering
		\includegraphics[width=\textwidth]{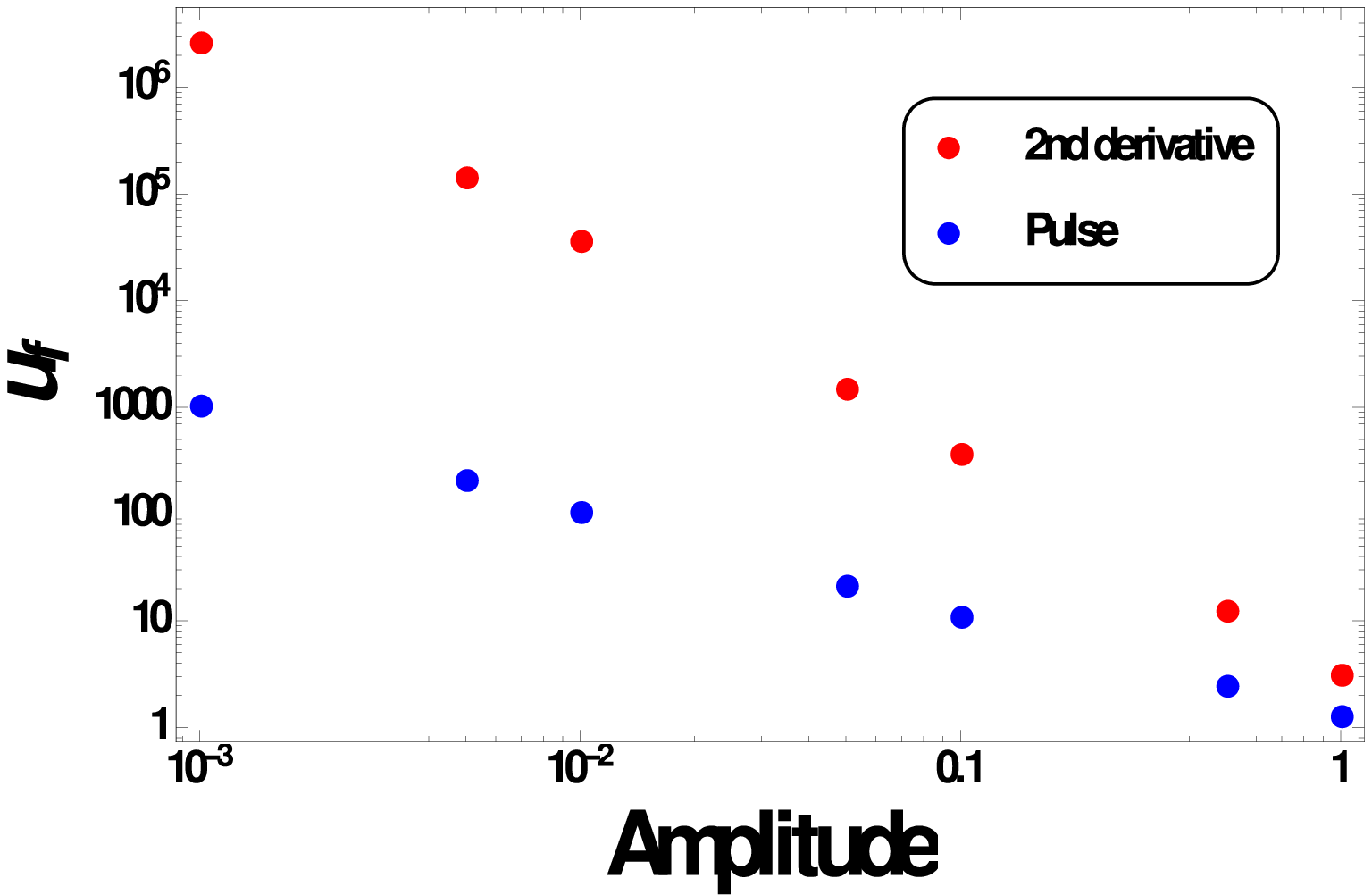}
		 \caption{}
		\label{fig:a_vs_uf_pulse}
	\end{subfigure}
	\caption{Plot of amplitude versus $u_f$ (focusing value) using logarithmic scales along both the axes. (a) $u_f$-amplitude plots for 1st and 3rd derivatives (odd functions) of
	a sech-squared pulse (odd function) (b) $u_f$-amplitude plots for the pulse itself and its 
	2nd derivative (even function). All values here are obtained by numerically solving Eqs. (\ref{eq:theta}) and (\ref{eq:sigma_plus}), with initial data at $u_i=-5$ where $\theta(u_i=-5)=\sigma_+(u_i=-5)=0$.}
	\label{fig:lensing}
\end{figure}

\noindent Figure \ref{fig:lensing} demonstrates more quantitatively this focusing nature of timelike congruences in the spacetime representing exact 
plane gravitational waves. In order to distinguish between effects for the pulse or its various derivatives, 
we have shown in Figs. \ref{fig:a_vs_uf} and \ref{fig:a_vs_uf_pulse}, the variation of $u_f$ (value of the affine parameter $u$ where focusing occurs) with the amplitude of the pulse\footnote{Amplitude here denotes the overall coefficient which appear in the functional expressions for a pulse or its derivatives. For example, $A_+(u)= a\sech^2(u)$ in the case of the pulse, where $a$ represents the amplitude. In Figs.(\ref{fig:a_vs_uf}) and (\ref{fig:a_vs_uf_pulse})  '$a$' ranges from $10^{-3}$ to $1$ (in arbitrary units)} for all cases studied here. We note that as the amplitude increases, focusing occurs earlier (lower values of $u_f$) and vice versa. Physically, it shows that the focusing 
value $u_f$ is dependent on the peak amplitude of the gravitational wave pulse
(or its derivatives) which seems to act like a converging lens. We plot $u_f$ versus the
amplitude for the odd functions [first and third derivatives, Fig. \ref{fig:a_vs_uf}] and even functions [second derivative and the pulse itself, Fig. \ref{fig:a_vs_uf_pulse}] separately. From 
the $u_f$ values shown in all the data displayed in Fig. \ref{fig:lensing}, we may conclude that focusing for the pulse happens earlier as compared to all other derivatives. In Fig. \ref{fig:a_vs_uf} we find that the third derivative focuses earlier in comparison to the first derivative and, with increasing amplitude, the difference in $u_f$ value
for the first and third derivative cases remains almost the same. In contrast, Fig. \ref{fig:a_vs_uf_pulse} 
suggests that the difference in $u_f$ values for the
pulse and its second derivative decreases with increase in amplitude. Therefore, in some sense, the $u_f$ versus amplitude plots may provide a way to quantify and differentiate between the effects due to a pulse and its various derivatives. 
As stated earlier, since the derivatives are linked to physical scenarios,
one may consider the $u_f$ versus amplitude plots as a way to quantify and
distinguish between the memory effects arising
in such contexts.

\noindent Finally, if we have an {\em initially expanding congruence} (unlike the ones
discussed above where we looked at an initially parallel congruence),
we have checked (not shown here) that focusing occurs and is dependent on the gravitational wave amplitude. Similarly if we have an {\em initially converging congruence}, there is always focusing.
Thus, for all types of initial configurations, we observe focusing as well as a permanent distortion (shear) for timelike geodesic congruences encountering a localized gravitational wave pulse (or its derivatives). 

\section{Conclusions}
\noindent We have, in this article, tried to arrive at an understanding of a memory effect using the kinematic variables that define a geodesic congruence, namely the expansion, shear and rotation. In the exact, vacuum plane parallel gravitational wave line element, the Raychaudhuri equations for timelike geodesic congruences have been written down and solved. We have exploited
the known fact that geodesic motion in such spacetimes
reduces to a motion in an effective two-dimensional ($x,y$) 
mechanical system where the coordinate $u$ acts
like time. The "geodesic Lagrangian" [when $A_\times (u)=0$] becomes that of
a nonrelativistic oscillator along $x$
and an inverted oscillator along $y$, with time ($u$)-dependent frequency \cite{zhang}. 
The equation for the coordinate $V(u)$ is redundant since its geodesic equation reduces to an identity. Following standard methods, we have obtained the behavior of $\theta$ and $\sigma_+ (\sigma_\times)$ for plus (cross) polarization for general as well as specific choices of the pulse profiles. There is no rotation ($\omega$) involved in the congruences we have worked with here. 

\noindent In contrast to noting a memory effect through geodesics or geodesic deviation, we have
 shown how kinematic variables like expansion and shear can carry information about memory. Qualitative treatment of the case 
 of a generic pulse or its derivatives lead to constraints on the values of expansion and shear as $u\to u_f$ (focusing). 
 These have been discussed in detail. 

\noindent Quantitative solutions (for specific pulses) 
for the expansion and shear obtained above are in full agreement with the inequality constraints found in the qualitative analysis. The plots for expansion and shear for both the pulse and its derivatives show divergences at specific values of $u=u_f$.
The exact location of $u_f$, expectedly, depends on the functional forms representing the pulse or its derivatives. 

 \noindent Furthermore, we note that the value of $u$ where
focusing happens (along with the growth/decay of shear) after the pulse has departed, depends on the amplitude and the width of the pulse.
 We have plotted $u_f$ as a function of the amplitude of the pulse. We observe that as the amplitude increases the value of $u_f$ shrinks, showing that the pulse focuses more acutely. Even for an initial expanding congruence, there is focusing dependent on the pulse amplitude thereby resembling what is seen in a converging lens. Considering this analogy with geometric optics \cite{Harte}, we may also 
 associate the amplitude of a 
 pulse with the inverse of the focal length. Hence, pulse-induced focusing coupled with a change of shear can act as yardsticks for understanding ${\cal B}$-memory.
 
 \noindent The fact that null geodesics do form caustics in exact, plane gravitational wave spacetimes of fixed polarization had been shown many years ago in the
 work of Bondi and Pirani\cite{bondi}. However, their work does not involve studying
 the behavior of the kinematic variables of geodesic congruences in order to
 figure out focusing effects or a change in shear. Further, in our studies, we demonstrate how such {\em benign focusing} for timelike geodesic congruences occur as induced by the appearance of a pulse. There is no real singularity in the spacetime (the invariant scalars are zero everywhere). The point of intersection of the geodesics implies deviation going to zero (which coincides with the location in $u$ where the expansion $\theta$ diverges to negative infinity) and is thus a critical point where the coordinate singularity of the metric (when written in BJR coordinates) appears \cite{zhang1}.
 Our explicit and detailed analysis of shear-induced focusing and its association with  
${\cal B}$-memory are both new and so are the numerous exact analytical
solutions showcasing the ${\cal B}$-memory effect directly.
 
 \noindent A more complete and detailed treatment of the Raychaudhuri equations with both the $+$ and $\times$ profiles simultaneously present can be an extension of our work.  It will also be interesting to note if rotation ($\omega$), when initially present, has any role to play
 in controlling the eventual focusing of geodesics. It is possible that rotation may
 prohibit focusing leading to finite changes in the expansion and shear.
 Studying the influence of a gravitational wave pulse on the evolution of the full ${\cal B}$-matrix can also be an elegant and unified approach 
 toward arriving at an associated memory effect. We conclude with the hope that in the future, such studies on the 
 kinematic evolution of geodesic congruences in relevant spacetimes of interest will be able to throw more light on 
 newer aspects of ${\cal B}$-memory. 
 
 \section*{ACKNOWLEDGMENTS}
 \noindent I. C. acknowledges the University Grants Commission (UGC) of the Government of India for providing financial assistance through senior research fellowship (SRF) with reference ID: 523711

\end{document}